# Low-coherence digital holographic microscope with Fizeau interferometer

MOHIT RATHOR , SHIVAM KUMAR CHAUBEY, NEHA CHOUDHARY, RAKESH KUMAR SINGH*

*Laboratory of Information Photonics and Optical Metrology, Department of Physics, Indian Institute of Technology (Banaras Hindu University), Varanasi, 221005, Uttar Pradesh, India*
*Corresponding author: krakeshsingh.phy@iitbhu.ac.in

We present a new digital holographic microscope (DHM) with a low coherent source for the quantitative imaging of smooth and optically rough objects. The experimental design of the microscope uses an in-line experimental geometry based on the Fizeau interferometer and shows the depth sectioning capability due to the limited longitudinal coherence of the source. A polarization-phase shifting approach is implemented to extract the quantitative and speckle-free image from the experimentally recorded interference fringes. To test and experimentally demonstrate the working of the proposed DHM, we present the results of the quantitative images for different objects.

Digital holographic microscope (DHM) is an important imaging technique with numerous wide ranges of applications. The DHM system is comprised of an optical design to experimentally record the hologram by a digital detector and then numerically reconstruct the digitally recorded hologram for quantitative phase imaging (QPI). With the numerical reconstruction of the hologram, the quantitative information of the object is readily available in the digital form for different applications. Hologram in the DHM usually results from coherent interference of the object and reference beams. Various DHM schemes have been proposed, and significant among them are in-line, off-axis, and phase-shifting holography [1,2]. The off-axis holography, which requires an angular separation between the object and reference beam to overcome the twin image issue of the in-line holography, is commonly used in the DHM. However, the angular separation between the object and reference beams may introduce differential environmental noise and external disturbances which results in stability issues. In order to overcome such challenges in the DH, in-line and common path configurations have been developed [3-6].

Integration of the phase shifting in the holography removes zero-order and conjugate images from the reconstruction of the off-axis hologram, hence utilizing the full bandwidth of the detector. The phase shifting is achieved through different methods such as a computer-controlled piezoelectric transducer [7], acousto-optic modulator [8], spatial light modulator [9], parallel phase shifts by polarization camera [10], and polarization phase shift [11]. The Fizeau interferometer with polarization phase-shifting offers an alternative and stable experimental method for QPI [12,13]. This experimental configuration offers an advantage over the commonly used interferometric geometries such as the Michelson or Mach-Zehnder. Nevertheless, holographic image quality is severely disturbed by the coherent noise due to the use of a spatially and temporally coherent light source. For instance, a reflection of a coherent beam from the optically rough object generates an undesirable speckle pattern due to the randomly scattered wavefronts and thus introducing unwanted artifacts in the reconstructed results [14,15]. Therefore, unique features of the DHM are suppressed in the speckle noise.

Many methods have been developed to remove the speckle noise such as incoherent summation [16,17], shifting and rotating objects [18,19], laterally shifting hologram aperture [20], and computational denoising [21]. A broad discussion on speckle reduction in the DHM can be found in Ref. [14] Besides speckle noise reduction by incoherent summation at the imaging plane, tailoring coherence of the source brings a different dimension and advantages in the DHM. The influence of the spectral bandwidth upon the temporal coherence of the beam is utilized in interferometry for tomography and profilometry [22,23]. The interference techniques can also be used to image through diffusive media by filtering the ballistic photons [24]. The coherence between two longitudinally separated points can also be tuned by the extent of a monochromatic incoherent planar source [25,26]. Rosen and Takeda used longitudinal spatial coherence for the surface profilometer [27]. Synthesis of the spatial coherence has been studied to develop different imaging techniques [28,29], and in the design of exotic light sources [26,30]. Other imaging schemes with low-coherent sources are depth sectioning [31], imaging self-luminous objects [32], optical coherence microscopy [33], and digital holography [34,35]. However, interference with low coherent sources demands a meticulous alignment of the interfering beams. Therefore, design of a compact and a highly stable experimental setup for the holography with low coherent source is an important requirement.

In this letter, we propose and experimentally demonstrate a new low-coherence DHM for quantitative imaging. This technique uses a narrow longitudinal spatial coherence for depth sectioning, and separation between the two interfering beams is adjusted in the Fizeau geometry. The unique feature of the proposed DHM lies in its ability to tune the path length difference in the interfering beams in an in-line geometry. Moreover, this experimental system overcomes issues like speckle noise, alignment, and stability of the typical DHM system. In order to use the on-axis Fizeau geometry, we use polarization distinguishability in the incident beam where one of the orthogonal polarization components of the light, say horizontal, is loaded with the object, and other polarization components, say vertical, works as a reference beam. A longitudinal separation between the orthogonal polarization components is kept within the coherence length of the source and thus permits desired interference and optical sectioning. Further, we apply polarization-phase shifting to digitally extract the complex field from the phase-shifted interference patterns. The polarization phase shifting is introduced between the orthogonal polarization components of the beam by a combination of a quarter-wave plate (QWP) and a polarizer. Details of the experimental designs, implementation, and results are discussed below.

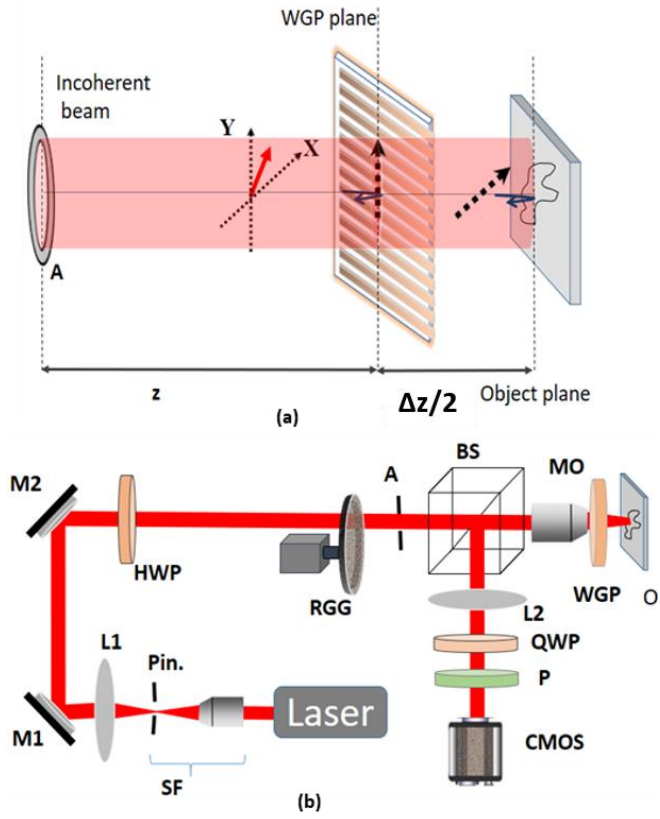

Fig. 1. Schematic diagram of the working of grid polarizer and experimental setup of the system, (a) Represents the working and a ray diagram at the wire grid polarizer, A is the aperture at the incoherent source plane, z is the distance between aperture to grid polarizer and (Δz/2) is the distance between grid polarizer plane to object plane, Orientation of the electric field vectors of the light are represented by X and Y directions; (b) Experimental setup, Pin. pinhole; L1,L2 lens; M1,M2 mirror; HWP half wave plate; RGG rotating ground glass; A aperture; BS beam splitter; MO microscopic Objective, WGP wire grid polarizer; O object; QWP quarter wave plate; P polarizer; CMOS Complementary metal oxide semiconductor camera.

The experimental design is shown in Fig. 1. A horizontally polarized He-Ne laser at $\lambda = 632.8\ nm$ (Thorlabs, HNL150LB) is filtered and collimated by a spatial filter assembly and a lens L1. Polarization of the laser beam is tuned to the diagonal state by a half-wave plate (HWP) and the beam is directed towards the rotating ground glass (RGG) to mimic a thermal light source. The experimental system is composed of two major parts. First is the generation of a low coherent source and control of the longitudinal spatial coherence. This is realized by a RGG and a circular aperture A at the RGG plane as shown in Fig. 1(a). A wire grid polarizer (WGP) reflects the vertical polarization component of the light coming from the RGG and transmits the horizontal polarization component towards the object. This geometry makes a tunable longitudinal separation $\Delta z$ between the two orthogonal polarization components. The second part of the experimental system is a combination of the Fizeau interferometer with a polarization phase shifting scheme as represented in Fig. 1(b). A compact DHM is facilitated by the WGP. The WGP (Thorlabs; WP25M-VIS) reflects the vertical polarization component from its plane and the beam is folded and directed by the beam splitter (BS) toward the detector. This polarization component acts as a reference beam. On the other hand, a transmitted horizontal polarized component from the WGP interacts with the reflective object and is subsequently directed toward the detector through WGP and the BS. This back-reflected horizontally polarized beam is imaged at the detector plane by an imaging system composed of a microscope objective (MO) with magnification (10X) of numerical aperture (NA)= 0.25 and a bi-convex lens. Finally, the object beam interferes with a reference beam using a combination of polarization optics such as QWP and polarizer (P). The interference pattern is recorded by a complementary metal-oxide semiconductor (CMOS) camera (Thorlabs, DCC3240M). Specifications of the CMOS are: pixel size 5.3 μm and pixels number 1280x1024.

A single realization of the light field immediately behind the RGG is described as

$$U_p(\rho, t) = |U_p| \exp[i\ \phi(\rho, t)], \qquad (1)$$

where $|U_p|$ is the amplitude of the uniform incident wave and $\phi(\rho, t)$ is the random phase introduced by a non-birefringent diffuser at a particular instant of time $t$. $p = x, y$ represents the orthogonal polarization components of the incident diagonally polarized beam. The random phase $\phi(\rho, t)$ is uniformly distributed over the interval $[-\pi, \pi]$ [36]. For a monochromatic beam, we suppress time $t$ in our notations. The corresponding realization of the complex field at the longitudinal distance $z$ is represented as

$$U_i(r, z) = \int U_i(\rho) G(r - \rho) d\rho, \qquad (2)$$

where $G(r - \rho)$ is the Fresnel propagation kernel and $r$ is the spatial coordinate at the observation plane.

A statistical property of the light can be characterized by second-order correlation $\langle U^*(r,z) U(r, z + \Delta z)\rangle$, where the angular bracket $\langle\ \rangle$ represents the ensemble average. Based on our experimental design, the longitudinal spatial coherence $\Gamma(\Delta z) = \langle U_y^*(r,z) U_x(r, z + \Delta z)\rangle$ is at peak within a narrow window of $\Delta z = 0.5\ mm$ for a circular aperture A= 3 mm. This permits interference fringes within this range and consequently depth sectioning in the DHM. The WGP is placed at a distance z=70 mm from the RGG plane, and separation between WGP and object is taken as ($\Delta z/2$)= 6.75 mm. A detailed discussion on tuning the longitudinal spatial coherence with source and angular radius can be found in Ref. [37,39]

The orthogonal polarization components, reflected and transmitted beam from the WGP, are represented as

$$U_y(r) = |U_y(r)|exp(i\phi_y(r)), \quad (3)$$

$$U_x(r) = |U_x(r)|exp(i(\phi_y(r) + \phi_0)), \quad (4)$$

where $|U_y(r)|$ and $|U_x(r)| = |O(r)||U_y(r)|$ are amplitude distributions of the reference and object-mediated waves, respectively. The phase distributions of the reference and object waves are represented as $\phi_y(r)$ and $\phi_x(r) = \phi_y(r) + \phi_0$, respectively. For the experimental configuration in Fig. 1(b), the vertical and horizontal polarization components have difference in the radius of curvatures due to difference in the reflection conditions of both the beams. Here, vertical polarization component is converged by the MO but immediately reflected at the WGP plane. On the other hand, the horizontal polarization component, as shown in Fig. 1(a), propagates to the object plane and then imaged at the camera plane.

To realize the quantitative phase imaging in an on-axis geometry, we use a polarization-phase shifting by the QWP and a polarizer combination. The fast axis of the QWP is oriented at 45° with respect to the horizontal axis, thereby converting the orthogonal polarization (horizontal and vertical) state into orthogonal circular polarization states. A polarizer (P) is rotated to four different orientations, namely 0°, 45°, 90° and 135° to record the four interference fringes at the CMOS plane. A realization of the beam after a QWP and a polarizer is represented as

$$\begin{pmatrix}U'_x(r)\\U'_y(r)\end{pmatrix} = \frac{1}{\sqrt{2}}\begin{pmatrix}\cos^2\theta & \sin\theta\cos\theta\\\cos\theta\sin\theta & \sin^2\theta\end{pmatrix}\begin{pmatrix}1 & i\\i & 1\end{pmatrix}\begin{pmatrix}U_x(r)\\U_y(r)\end{pmatrix} \quad (5)$$

Eq. (5) represents the Jones matrix corresponding to the combination of a QWP and a polarizer oriented at an angle $\theta$. Therefore, the averaged intensity distribution at the detector plane is given as

$$I(r,\alpha) = \langle(U'^*_x(r) \quad U'^*_y(r))\begin{pmatrix}U'_x(r)\\U'_y(r)\end{pmatrix}\rangle \quad (6)$$

$$I(r,\alpha) \approx 1/2\,[\langle U_x^2(r)\rangle + \langle U_y^2(r)\rangle + \langle 2\,U_x(r)U_y(r)\sin(2\theta + \phi_0(r))\rangle], \quad (7)$$

where $\langle\;\rangle$ is replaced by summation over the independent realization of the intensities due to the RGG, $\alpha = 2\theta\,(0,\frac{\pi}{2},\pi,\frac{3\pi}{2})$ is the phase shift between reference and object beams and $\phi_0(r)$ is the phase distribution of the object field.

The complex amplitude distribution of the object field is obtained using the 4-phase shifted interferograms as [11]

$$O(r) = [I(r,0) - I(r,\pi)] + i\left[I\left(r,\frac{3\pi}{2}\right) - I\left(r,\frac{\pi}{2}\right)\right], \quad (8)$$

where $O(r) = |O(r)|\exp[i\phi_o(r)]$ is the imaged complex field of the object Customized longitudinal spatial coherence support depth sectioning and different depths of the objects are imaged by mechanically moving the imaging camera. The object is placed at the longitudinal plane $z + \Delta z/2$. We consider two different objects for imaging and the results are presented in Figs.2 and 3.

In Fig. 2, (a)-(d) show the recorded on-axis intensity holograms for different phase shifts ($0,\pi/2,\pi,3\pi/2$) between orthogonally polarized reference and object beam. Here, we have used focal length 200 for lens L2 and the object is introduced in the beam by a reflective type SLM (Pluto 2.1, Holoeye). This SLM introduces a phase structure in the horizontally polarized beam. Figs. 2(a)-(d) represent the interferogram with 255 gray level at the SLM. On the other hand, Figs. 2 (e) –(h) represent phase-shifted interferograms for the phase object $\beta$ displayed by the SLM. Figs. 2 (i) and (j) show the reconstructed quantitative amplitude and phase information by a four-step phase shifting approach for blank screen (without object and gray level 255) of the SLM. Spherical rings in the phase profile in Fig. 2 (j) appear due to experimental configuration where the reference beam is immediately reflected from the WGP and the object beam is imaged at the CMOS plane, and this result is used for calibration to cancel residual phase variations in the experimental system. On the other hand, Figs. 2 (k) and (l) show reconstructed quantitative amplitude and phase information by a four-step phase shifting approach for the object $\beta$ which is calibrated with respect to results of the blank srcen SLM.

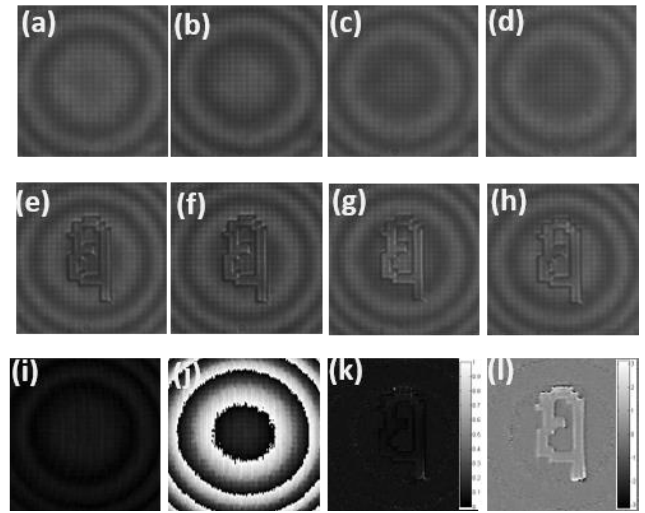

Fig.2. Experimental results (a)-(d) is the phase-shifted holograms for the blank screen(grayscale 255) on SLM, (e)-(h) is the phase-shifted holograms for the object $\boldsymbol{\beta}$, (i) & (j) is the amplitude and phase information of the blank screen and (k) &

(l ) is the amplitude and phase information of the object $\beta$ after the calibration with blank screen.

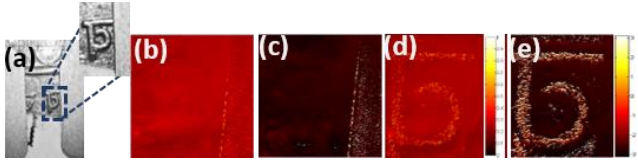

Fig.3. Experimental results (a) original optically rough object (coin with a blade); Results with developed low coherence DHM (b) & (c) is the amplitude and phase information of blade, wherein coin is out of focus, (d) & (e) is the amplitude and phase information of the coin where blad is out of focus due to depth sectioning.

Fig.3 shows the performance of the system with an optically rough object. Here, focal length of the lens L2 is 100 mm and other specifications of the imaging system are the same as in the previous case. Imaging of such an optically rough object with a coherent light source is a challenge due to dominant speckle pattern. Here, we overcome this issue and demonstrate the use of our low-coherence DM method. We imaged a layered rough object which has two rough surfaces one is a coin and another is a blade (placed in front of the coin) as shown in Fig. 3 (a). Due to the limited longitudinal coherence of the source in the DHM, we are able to quantitatively image only one plane at a time from the interference patterns. In order to image the second plane, we mechanically move the object plane to bring another plane in the coherence window.  First, we image the blade using a translation stage and record the four on-axis intensity holograms with phase shifts, reconstruct the quantitative amplitude and phase information of the blade surface as shown in Fig.3 (b) and (d) respectively. Second, we move the focus on the coin surface and record the phase-shifted interferograms. The quantitative amplitude and phase information of the coin surface is shown in Fig.3 (d) and (e) respectively.

In conclusion, we present a new on-axis digital holographic microscope for quantitative imaging of optically rough objects and avoid the challenge of laser speckle. This method overcomes the issue of speckle noise in quantitative imaging by reducing the coherence of the light source rather than relying on speckle averaging at the observation plane. The experimental setup brings a highly stable configuration for digital holography with the Fizeau-interferometer. The possibility of tailoring the longitudinal coherence of the source and on-axis tuning of the longitudinal separation between the interfering beams brings a depth sectioning capability to this imaging system.  Quantitative phase and amplitude information of the object is reconstructed by using the principles of polarization phase shifting and using four such phase-shifted interferograms. Experimental tests and results presented as a demonstration of the proposed technique highlight its potential for quantitative imaging of optically rough objects and for extending the scope of the DHM.

**Funding.** Department of Biotechnology (DBT) grant no. BT/PR/35587/MED/32/707/2019; Science and Engineering Research Board (SERB)- CORE/2019/000026.

**Acknowledgment**. Mohit Rathor acknowledges fellowship from the IIT (BHU).